# The Effect of Thermal History on Microstructural Evolution, Cold-Work Refinement and α/β Growth in Ti-6Al-4V Wire + Arc AM

J. R. Hönnige[1], P. A. Colegrove[1], P. Prangnell[2], A. Ho[2], S. Williams[1]

[1]Welding Engineering and Laser Processing Centre, Cranfield University, MK43 0AL Cranfield, UK

[2]School of Materials, Manchester University, Grosvenor Street, Manchester M1 7HS, UK

## Abstract

Wire + arc additive manufacture (WAAM) is an attractive method for manufacturing large-scale aerospace components, however the microstructural changes that occur and the effect of inter-pass rolling are poorly understood. Therefore two fundamental studies were conducted: the first involved temperature measurement of a wrought "dummy" wall so that the microstructural changes in the heat affected zone (HAZ) could be related to the thermal cycle. This demonstrated that the white band in the microstructure corresponded to 825 °C – well below the β-transus temperature – and above this boundary the bi-modal substrate material was converted to lamellar. The second involved peening WAAM material along the side of a deposited wall before applying a typical WAAM thermal heat treatment. This showed that refinement occurred up to the first layer band in the microstructure and the smallest grains were observed just above this boundary – at higher temperatures significant grain growth occurred. This study has provided the foundational understanding of microstructural changes that will facilitate future process developments.

## 1 Introduction

### 1.1 WAAM

There are a variety of high-deposition-rate additive manufacturing (AM) techniques, with one of the most cost-effective being Wire + Arc Additive Manufacturing (WAAM) [1]. WAAM targets the manufacture of near-net shape parts of meter-scale structural components with medium complexity, which are otherwise produced by forging, casting or machining from a solid block. It can significantly reduce the lead time, production cost and the material usage, while having competitive mechanical properties [1], [2]. Especially the production of titanium components exploits the advantages of WAAM due to commonly high material and process costs. The mechanical performance of additively manufactured titanium is attributed to its' typical microstructure.

## 1.2 Microstructure in Ti–6Al–4V WAAM

Figure 1 (a) shows a schematic of a typical Ti–6Al–4V WAAM microstructure of a cross-section in a single pass wall, in which the following typical landmarks can be found: the fusion boundary cannot be identified easily with optical microscopy, but it is understood to be convex, unlike the concave shape typical for welding [3], [4]. The heat-affected zones (HAZ) are much more clearly viewable and appear as horizontal, isothermal lines, and are called "layer bands". There is some debate over what they represent with some authors stating that they represent the β-transus temperature of 995 °C [5], while others stating that it corresponds to the α-dissolution temperature [6], which is 748°C for Ti–6Al–4V [7]. Below the top layer band, the established microstructure constitutes repetitive layer bands, which are equidistant and equal to the layer height [8]. The size of the α plates increases from one layer band to the next, as shown in Figure 1 (a) [6]. For Ti–6Al–4V, the mechanical properties are mainly driven by the size of these α-plates in the lamellar microstructure. The diffusion distance during β→α transformation of the β-stabilising vanadium determine the α-plate size [9], as they form the retained β-phase, which separates the α-plates from each other. Their size is therefore a result of the transient cooling rate between the β-transus and the temperature, when the β→α transformation is completed [10], [11].

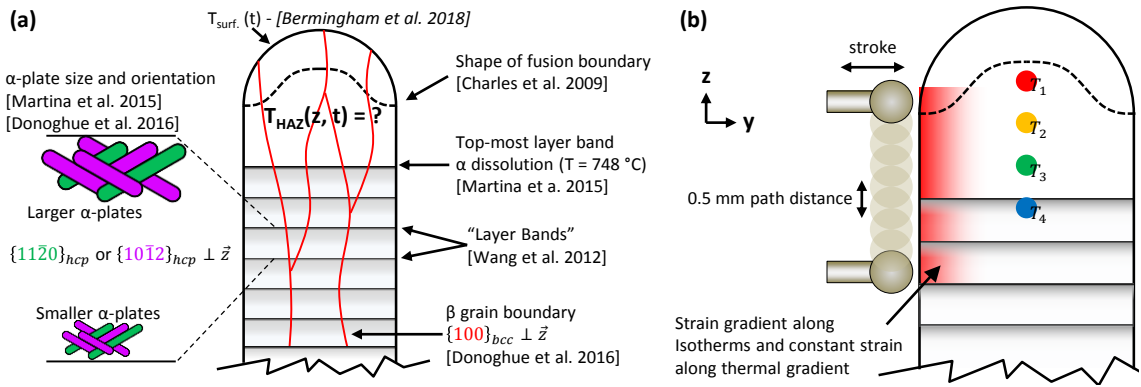

**Figure 1: (a) Schematic of a typical cross-sectional Ti–6Al–4V microstructure in single-bead deposition; (b) visual summary of the two experiments presented in this paper.**

Process parameters in AM strongly influence the cooling rates and therefore determine the properties in the as-built condition of (AM) components [12]–[14]. In addition, the mechanical properties are also affected by a strong texture. The texture is due to epitaxial growth and the development of large columnar β-grains with a strong dominance of the {100} fibre of the body-centred-cubic (bcc) unit cell in the build direction of the deposit [15]. During the subsequent β→α transformation the β-phase imparts its' texture on the α-morphology according Burgers orientation relationship (BOR) [16]. This results – depending on the variant selection – in the alignment of either the $\{1120\}_{hcp}$ fibre or the $\{1012\}_{hcp}$ fibre with the z-direction [15], [17] and ultimately in anisotropic mechanical properties [8], [18].

### 1.3 Thermal History Investigation

The thermal history that establishes this microstructure is located between the fusion boundary and the top-most layer band of a deposition (within the HAZ - Figure 1 (a)). A number of investigators have sought to understand the thermal history and the resultant microstructure in WAAM. Even though the simulation of the thermal history in the bulk is possible [3], temperature measurement is very challenging and has only been performed on the surface [6], [19]. Martina et al. [6] measured the temperature/time history using thermocouples at the layer band and found that the maximum cooling rate was 7.4-8.4 °C/s and the peak temperature at the band was 740°C and near the α-dissolution temperature of 748°C. However, it is the thermal history between this location and the fusion boundary that matters for the final microstructure and the associated mechanical properties. Birmingham at al. [19] recently used a calibrated and contactless IR pyrometer to measure the surface temperature behind the molten pool after solidification. They also discussed the effect of the high heat-input and lower cooling rates of WAAM (10-20 °C/s through the β→α transformation field) on the α/β-morphology, which results in lower strength, but higher ductility, compared to powder-bed AM [20]. Temperature measurement inside this region is extremely difficult [21]. A gradient of more than 100 °C/mm [19] makes precise thermocouple positioning essential. Positioning thermocouples on the surface near the molten pool would directly expose them to the zone of influence of the WAAM welding torch, which could falsify the measurement or destroy the thermocouple. This can be mitigated by placing them in drill holes [6]. Finally standard K-type thermocouples were found to be incompatible with the Ti–6Al–4V alloy at higher temperatures, and so the use of the costly R-type is required in the location of interest. A different approach is the use of thermal imaging [22], however experience has shown that this method is extremely unreliable due to non-linear changes in emissivity due to surface oxidation [23]–[25]. In addition, thermal imaging is difficult to apply to liquid, glowing metals, as the emissivity changes significantly.

### 1.4 Post-Processing and Inter-Pass Rolling

Thermal post-processing can be deployed to amend process-dependent properties. Undesired α' martensite in powder-bed AM for example, which forms due to very high cooling rates above 410 °C/s during β→α transformation [26], are addressed with aging or hot-isostatic pressing (HIP). HIPing of AM parts has the additional benefit of eliminating defects, such as gas pores or voids, which would also diminish fatigue properties, just like the brittle α' martensite [20]. Defects however are very uncommon in Ti–6Al–4V WAAM using plasma deposition and the material is already ductile [8], which is why HIPing has no beneficial effect on the properties [19]. Solution treatment and aging can produce improved α-morphology within the parent β-grains [19], but it does not influence the β-morphology itself. The β-grain size and whether a columnar or a equiaxed microstructure forms, depends only on the temperature gradient during solidification and the solidification speed [27]. Once established, the β-morphology cannot be refined subsequently using thermal post-processing.

Inter-pass cold work on the other hand effectively eliminate the texture in Ti–6Al–4V and refine both the α-plates and the β-grains. The benefits are dramatically improved (α morphology) and isotropic (β-morphology) mechanical properties [2], [6], [15]. Donoghue [28] suggest that rolling causes twinning of the residual β-phase surrounding the α-plates, even though it is understood to be less favourable in Ti–6Al–4V [29]. When reheat by the subsequent deposition, these twinned grains cannot grow coherent anymore, form individual grains and ultimately result in a refined β microstructure. Even though it was demonstrated that β-refinement also causes α-refinement, a conclusive explanation has not been made yet [6].

## 1.5 Summary

In this paper a fundamental study was performed first to precisely measure the thermal history with thermocouples at different locations in the HAZ. In the second part, plastic strain was induced uniformly to the side of WAAM wall using machine hammer peening (MHP) to reveal the effect of the identified thermal history on the refinement process. Both studies are illustrated in Figure 1(b).

## 2 Methodology

### 2.1 Temperature History in the HAZ

To investigate the relationship between the microstructural changes and the thermal history, a Ti–6Al–4V dummy wall was machined from a hot rolled Grade 5 plate (ASTM B265), annealed for 60 min at 840 °C with globular microstructure and the composition shown in Table 1. Wrought material was selected over WAAM material to ensure that the microstructure was the result of one thermal cycle only and not the accumulation of many thermal cycles.

Table 1: composition in wt% of the Ti–6Al–4V Grade 5 ASTM B265-10/AMS 4911L raw material

| Ti | Al | V | Fe | C | N | O | Y | H | res. |
|---|---|---|---|---|---|---|---|---|---|
| bal. | 6.02 | 4.1 | 0.14 | 0.01 | 0.02 | 0.15 | <0.005 | 0.003 | <0.4 |

This wall had the same dimensions as a typical WAAM single pass wall, being 6 mm wide and semi-circular on top. The wall was 15 mm high and 250 mm long. Two sets of four equidistant (14 mm apart) holes that were at varying depths (1.5 mm difference) were drilled from the bottom of the wall into the expected HAZ (Figure 1 (b)). R-type thermocouples were spot-welded to the tip of the hole. The side-section of the specimen in Figure 2 shows the locations.

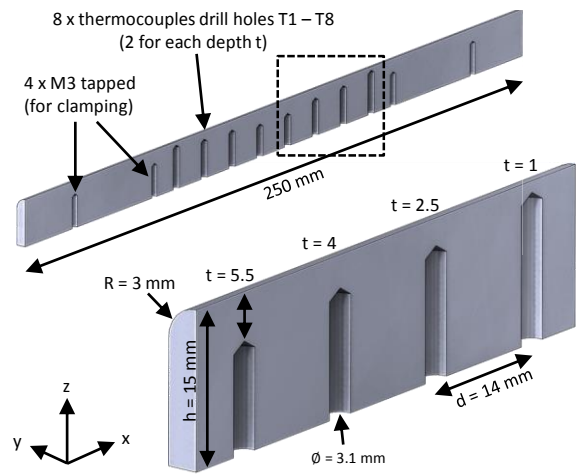

Figure 2: (a) The substrate: dimensions and thermocouple locations in the side-section view; (b) enlarged view of (a).

3.1 mm diameter holes were drilled to accommodate the 3 mm diameter ceramic thermocouple wire insulation. The wall was attached to a 300 x 60 x 7 mm aluminium plate with M3 bolts as shown in Figure 3 (a). A long slot hole in the centre of the aluminium plate allowed the thermocouples to be connected from the bottom (b). This setup was in turn mounted to a large aluminium backing bar. Titanium has poor thermal conductivity so this change to the overall structure, i.e with the insertion of the thermocouples and clamping to an aluminium substrate was not expected to significantly affect the thermal profile.

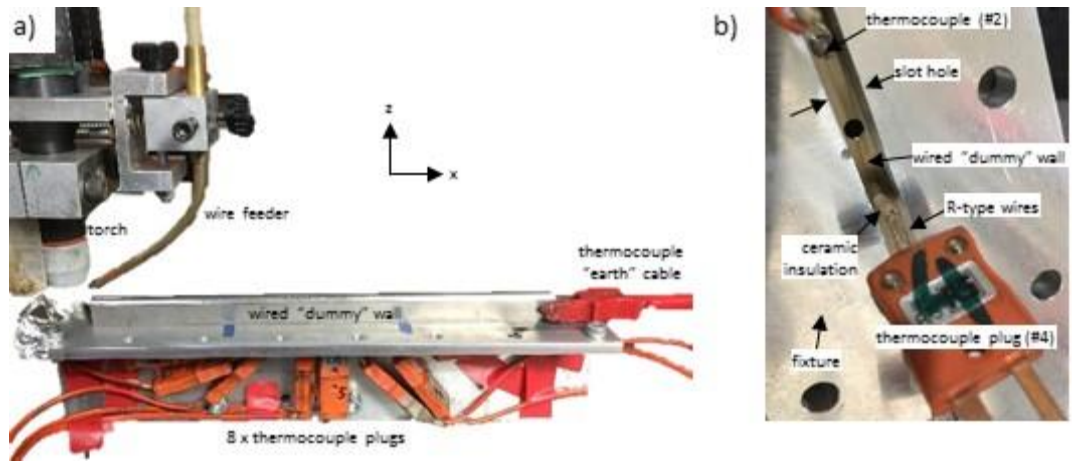

Figure 3: (a) Setup of experiment and (b) bottom view of the thermocouple connection.

An earth cable was attached to the wall to prevent interference between the thermocouple logger and the welding equipment. Ti–6Al–4V wire was deposited on the round edge, as if it was the next layer of a wall. An EWM-T352 tetrix plasma power supply was used with the deposition parameters from Table 2 which produced a 1.2 mm high layer. This experiment was repeated three times with three different walls, which each produced eight readings – two for each location. Wall three was then heat treated three times, by running the torch over the top layer without adding wire. The purpose was to quantify the thermocouple measurement error, as the thermocouples remained at the same location. Side sections and cross section cuts were performed on the thermocouple locations of wall 1 and 2 to investigate how the thermal history affected the raw material microstructure. The specimens were hot mounted in conductive resin, ground and polished. Scanning electron microscopy (SEM) was performed with 3500x magnification and 20 kV using the Back Scatter Detector (BSD). The SEM locations were located at the four thermocouple locations, the molten pool and the layer band. The microstructure was compared against globular microstructure of the substrate material shown in Figure 4.

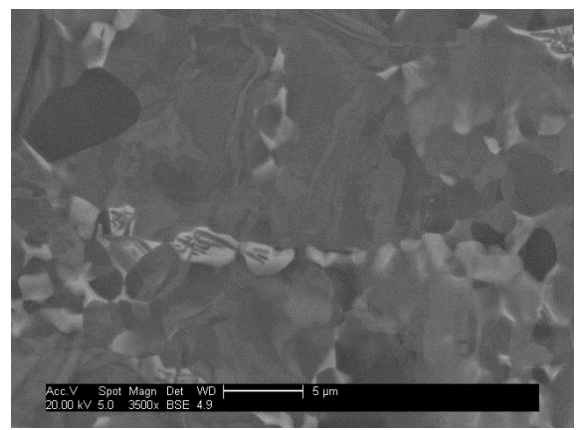

Figure 4: SEM image of the initial microstructure of the Ti–6Al–4V substrate (343 ± 3 HV)

Table 2: Welding Parameters

| Parameter | Value | Parameter | Value |
|---|---|---|---|
| Current [A] | 145 | Work Piece Distance [mm] | 8 |
| Wire Feed Speed [mm/s] | 40 | Plasma Gas Flow [l/min] | 1 |
| Travel Speed [mm/s] | 7 | Shielding Gas Flow [l/min] | 10 |
| Wire Diameter [mm] | 1.2 | Trailing Gas Flow [l/min] | 25 |

## 2.2 Machine Hammer Peening and Grain Refinement

To understand the effect of thermal history on grain refinement in a plastically deformed specimen, a WAAM sample was peened along the side and then a thermal heat treatment was applied with a plasma torch. The parameters from Table 2 were used again to deposit a 250 mm long wall on a 300 x 60 x 7 mm substrate which resulted in a 5.8 mm wide wall. After completing 15 layers, the wall had an established microstructure with large columnar prior β-grains [30]. Machine Hammer Peening (MHP) was applied with a ForgeFix XP tool using a 2 mm tip radius. It was applied to one side of the wall along its' entire length between the estimated location of the fusion boundary and the third layer band from the top – a width of 5 mm. This is shown in the schematic in Figure 1 (b) and in the image of the setup in Figure 5. It was designed to produce a uniform strain distribution in the z-direction along the side of the wall. The MHP travel speed was 36 mm/s in the x-direction and the tool's frequency was 180 Hz, producing one impact every 0.2 mm. Each of the nine following passes of the MHP tool were performed in alternating directions at a step distance of 0.5 mm.

Finally the plasma power supply was used again with the same deposition parameters, but this time without adding wire into the molten pool. The purpose was to apply the typical WAAM thermal cycle to the HAZ. The geometry of the wall did not change during this step. Two cross sections were taken in total, of which one was in the as-peened condition before the heat treatment and the other after. Wire was not added to ensure plastic deformation closer to the subsequent fusion boundary. Both cross sections were prepared for EBSD scanning, in an identical way to the previous study. EBSD scans were performed on both specimens in the top part of the wall. The EBSD data was then post-processed to reconstruct the prior β-grain orientation and to investigate the plastic deformation according to the method developed by Davies [31]. The method involves calculating most probable common parent β-orientation of neighbouring α-plates considering the possible variants during phase transformation respecting the BOR. Finally, the polished samples were etched with hydrofluoric acid to reveal the microstructure optically.

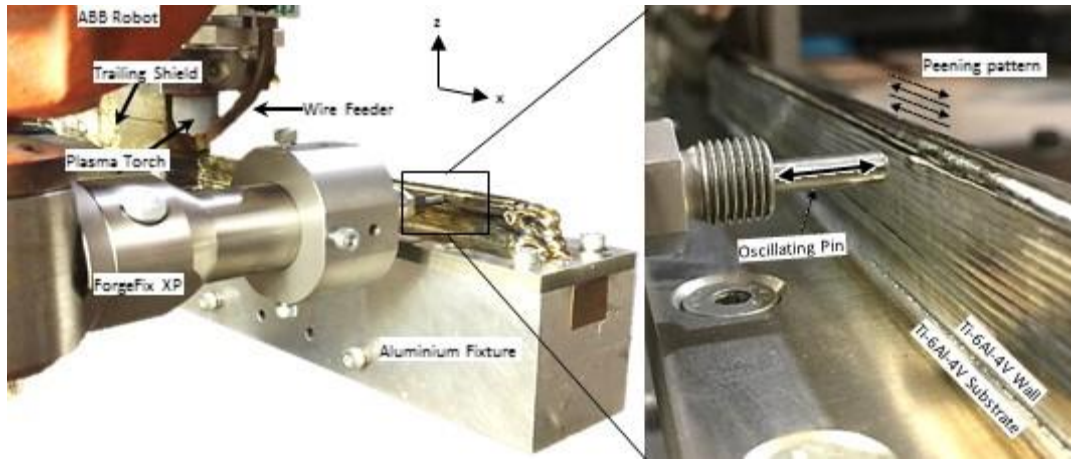

Figure 5: Side peening using the MHP tool ForgeFix XP.

# 3 Results

## 3.1 Thermal History

Figure 6 shows the peak temperatures of all thermocouple readings. Five of the 24 thermocouples failed during the experiment and the ones that did not fail were mostly in good agreement. The main anomaly occurred between the readings for thermocouple 1 and 8 even though their distance to the top of the wall was identical. The readings of all T8 thermocouples also showed a disproportional amount of noise. A measurement error due to interference with the welding current, when the torch was close to the thermocouple T8 and the "earth"-cable is suspected. The measurements where this was observed are therefore presented in brackets and excluded from consideration. Figure 7 (a) shows the thermal history within the HAZ from thermocouples 1 to 4 in wall 2. To aid comparison of the profiles, an offset of 2 sec was applied to the respective thermocouples which were 14 mm apart in the welding direction so they are presented as if the measurements were done at the same location.

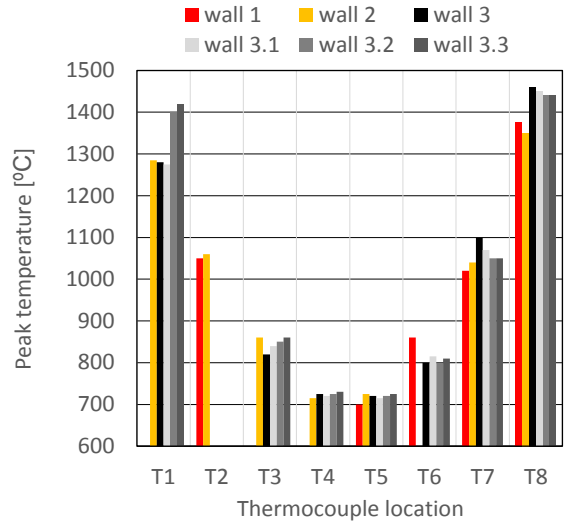

Figure 6: Peak temperatures of all thermocouple readings. The values of T1 and T8 that exceeded 1400 °C showed interference with the welding current and were disregarded.

Also indicated is the time that each thermocouple was above the β-transus temperature and how long it took to cool back down below the α-dissolution temperature. The actual time depended on heating and cooling rates derived from the thermocouple readings which are plotted in Figure 7 (b) and (c) respectively. All thermocouple 1 readings also showed a unique feature. It is difficult to identify due to the extremely high heating rates, but at about 1200°C, the heating curve seemed to bend slightly, which can also be identified with the heating rate profile Figure 7(b). Thermocouple 1 results from other specimens show this deviation slightly more clearly. Upon cooling this temperature profile had a short plateau at around 10 sec, where the cooling rates dropped from −20 °C/s to −10 °C/s, after which it increased again to −20 °C/s. This deviation also occurred with thermocouple 2 but is much harder to identify.

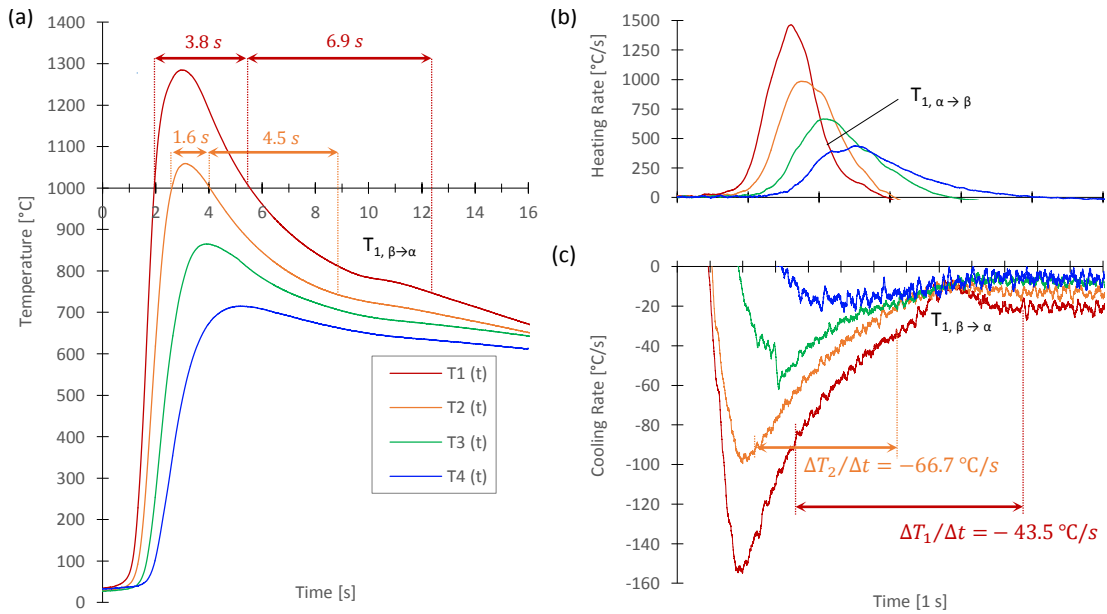

**Figure 7:** (a) Thermocouple reading of T1 – T4 in wall 2 and the derived (b) heating rates and (c) cooling rates

The average gradient of the peak temperature between thermocouple 1 and 4 as a function of depth below the top of the wall is 125 °C/mm. The bottom of the molten pool is approximately 0.7 mm above thermocouple 1, which highlights the sensitivity of this measurement to the position. A cross-section showing the thermocouple holes for location 1 to 4 with the associated microstructure is shown in Figure 8. The uniform heat input produced relatively constant grain morphology in the travel direction. At the location of the thermocouples, the horizontal features "dip" slightly downwards at thermocouple locations. The dip downwards makes sense due to a lack of material and corresponding heat sink. In any case the evidence suggests that the effect of the thermocouple holes on the thermal profiles is minor. During the thermocouple measurements, three errors can influence the measurement. The first is the systematic error of the temperature measurement, indicated by the vertical error bars. The second is another systematic error, in which the thermocouple drill hole increases the temperature at the measurement location. This would result in the measurement of a higher temperature or the temperature at a location closer to the heat source and is therefore represented by the left-hand side horizontal error bar. The right-hand side error bar shows potential inaccuracy of the thermocouple spot-weld location, which could have been applied not in the tip of the cone and therefore further away from the heat source.

Three hardness profiles were taken at two locations for each of the first two walls as shown in Figure 8 resulting in 12 measurements. 22 indents failed and were removed from a total of 588 measurements. The average profile with local standard deviation errors is shown in Figure 9 – the error in each measurement point was between 4 HV and 8 HV. The hardness in the unaffected substrate material was 332.9 ± 2.9 HV.

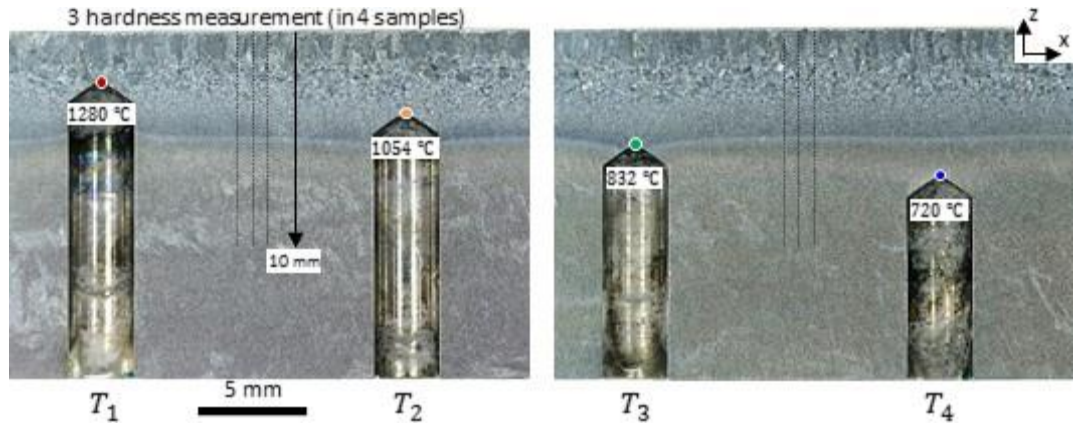

Figure 8: Side section of the microstructure at the thermocouple locations T1 – T4 and hardness profile locations

The hardness values can be compared against the microstructural features and the peak temperatures. Solidification of molten material, which is the 1.2 mm added material plus approximately 0.3 mm remelting resulted in the highest hardness. There is a visible transition from epitaxially grown columnar β grains to the equiaxed ones and a corresponding drop in hardness beyond this point.  A logarithmic trend line of the 5 temperature points was calculated, which enabled the temperature at this location to be determined.  It confirmed that the melting temperature (1600°C) was reached and the coefficient of determination ($R^2$) of the trendline was 0.9998. Beyond this point, the equiaxed β-grains are slightly softer, however the hardness increases slightly prior to the thin white band. Below this, the hardness decreases dramatically in the material whose peak temperature was between 600 °C and 850 °C. The microstructure in this wide region appeared slightly brighter. At distances greater than 8.5 mm from the top of the wall, where the temperature did not reach 600 °C, neither the hardness, nor the microstructure shows any change.

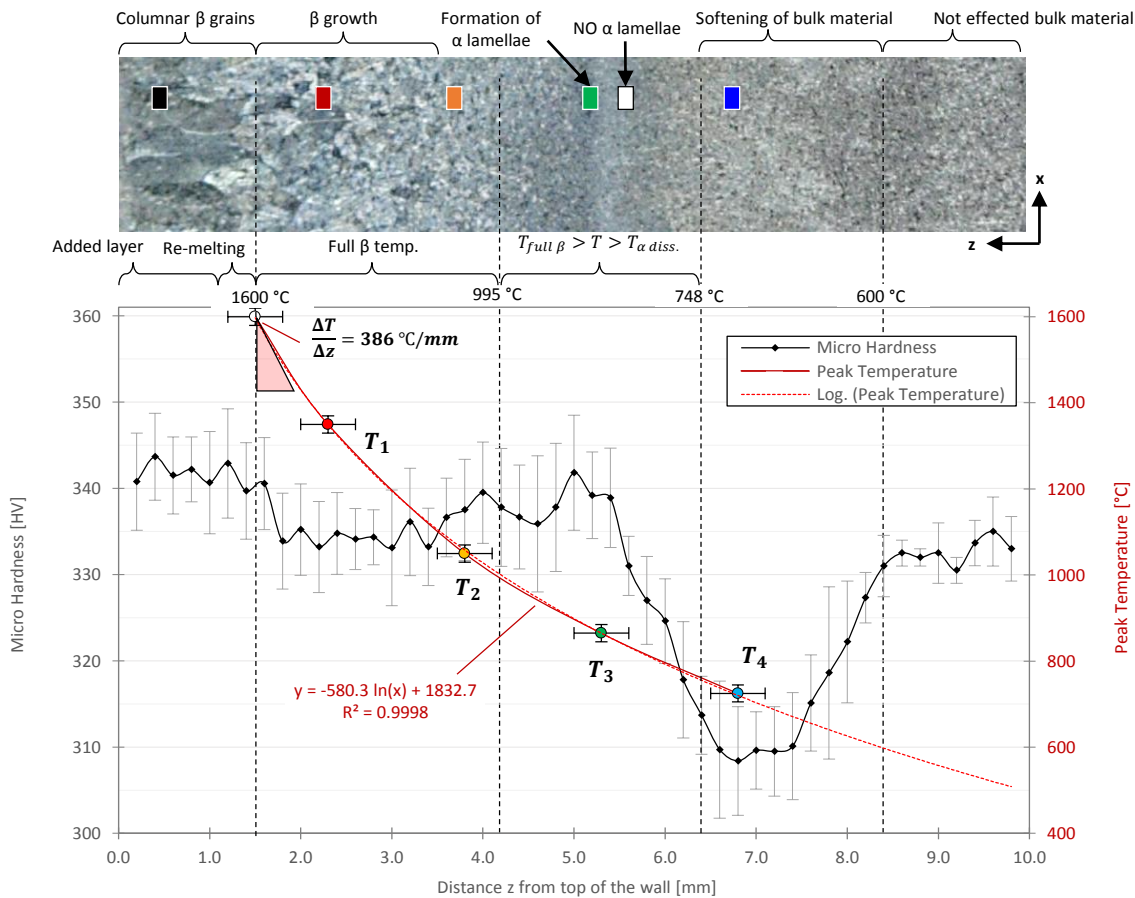

**Figure 9: Plot of hardness and temperature vs. distance from the top of the deposit with the corresponding microstructure.**

The SEM images, the locations of which are indicated in Figure 9, are shown in Figure 10 together with the respective hardness values and peak temperatures. In the solidified molten pool Widmanstätten α-plates of approximately 1 µm thickness were formed, some gathering in small colonies up to a few α-plates (a). Virtually no difference can be identified in the microstructure at the location of thermocouple 1 (b). The microstructure at thermocouple 2 (c) is similar, apart from a few colonies that consist of 5 to 10 very thin α-plates that are roughly 0.3 µm thick. The same thin α-plates can be found at thermocouple 3 (d) as well, but here they form relatively large colonies of 20 α-plates or more. Only 0.5 mm further away from this SEM location (e), at the location of the steep hardness gradient, can no α plates be found whatsoever. The estimated peak temperature was 825°C at this point was significantly higher than the α-dissolution temperature. The same microstructure can be observed at the thermocouple location 4 (f). Indeed the microstructures of both, Figure 10 (e) and (f), are not significantly different to the control microstructure of the substrate material in Figure 4, despite, although they exhibit blurred grain contours, referred to as "ghost microstructure" [32].

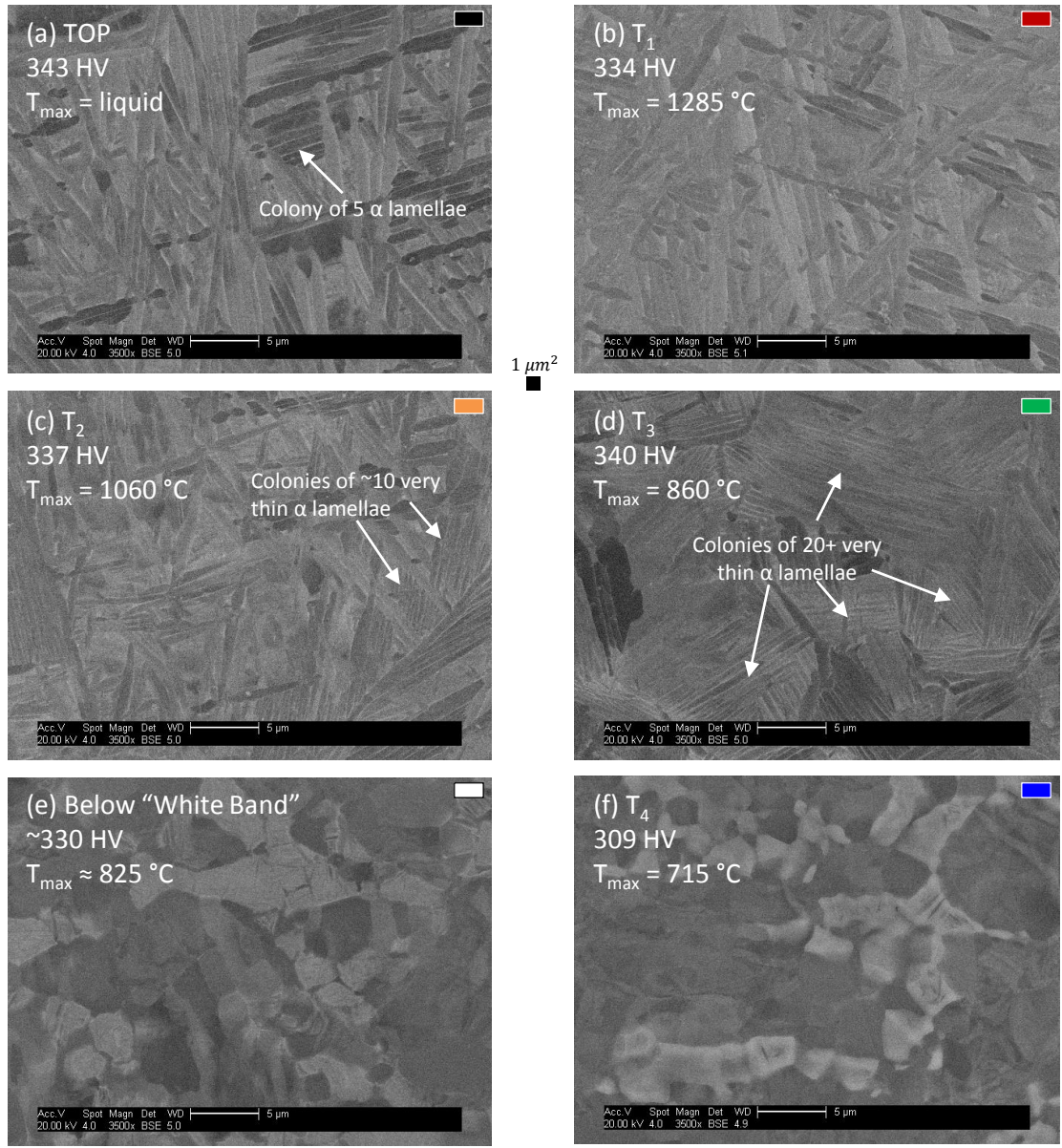

**Figure 10:** SEM Images from different locations: (a) inside the molten pool; (b) $T_1$; (c) $T_2$; (d) $T_3$; (e) below the white band and (f) $T_4$.

## 3.2 Grain refinement after MHP

Figure 11 (a) shows the microstructure in the as-deposited Ti−6Al−4V WAAM wall after machine hammer peening was performed on the left side of the wall. The dashed line represents the previous shape prior to deposition which was reproduced by offsetting the top of the wall by the layer band distance observed in the bottom part of the image.

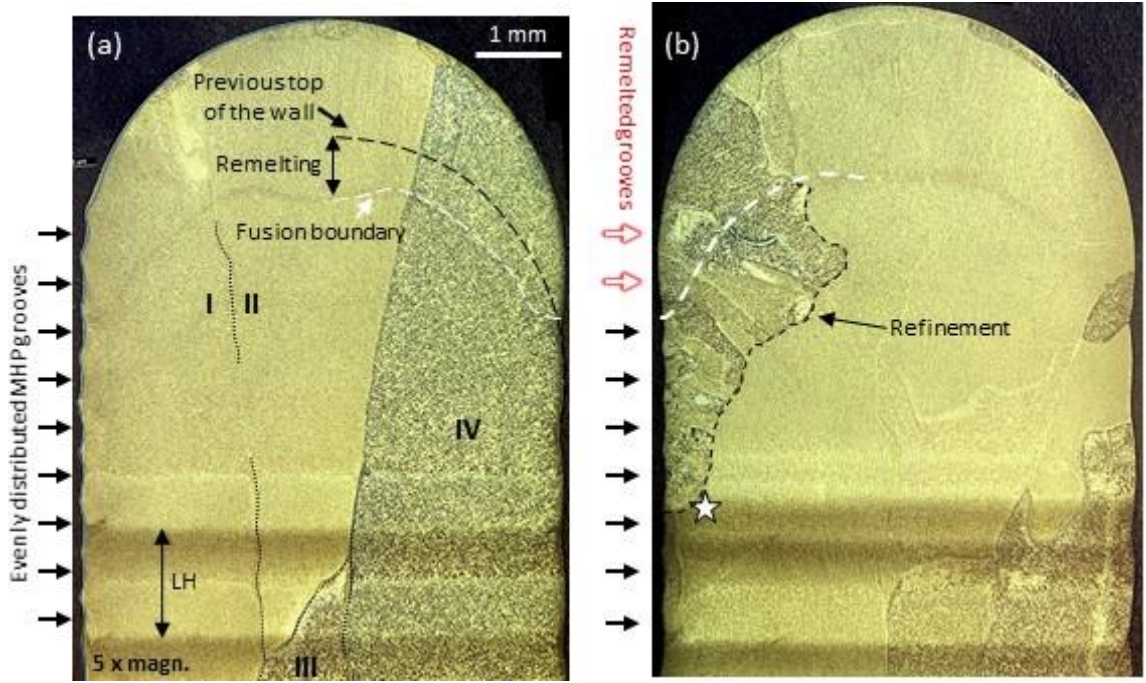

**Figure 11: Microstructure of the as deposited and as-peened**

A faint outline of the fusion boundary can be observed due to slightly darker shading below the previous top of the wall, indicated by the bright dashed line. The area in between is therefore the remelted material. Four individual large columnar β-grains were identified and are labelled with Roman numerals I – IV and they grew epitaxially through several layers. The equidistant indentations on the left side of the deposit are the MHP passes, which deformed the left side of the wall between top left corner and the second last layer band. The refined microstructure after the subsequent heat treatment with the torch is shown in Figure 11 (b). The MHP indents are still visible apart from the top two passes, which were completely remelted. No refinement was observed below the top most layer band, indicated with the star. The depth of the refinement at this location is nearly 1 mm deep, while close to the fusion boundary it reaches almost to the centre of the wall. The α-phase inverse-pole-figure (IPF-z) maps in respect to the z-fibre of both specimens are shown in Figure 12 (a) and (b). The dominant alignment of the hexagonal-close-packed (hcp) unit cell is with the magenta-coloured $\{10\bar{1}2\}_{hcp}$ plane normal to the build direction z.

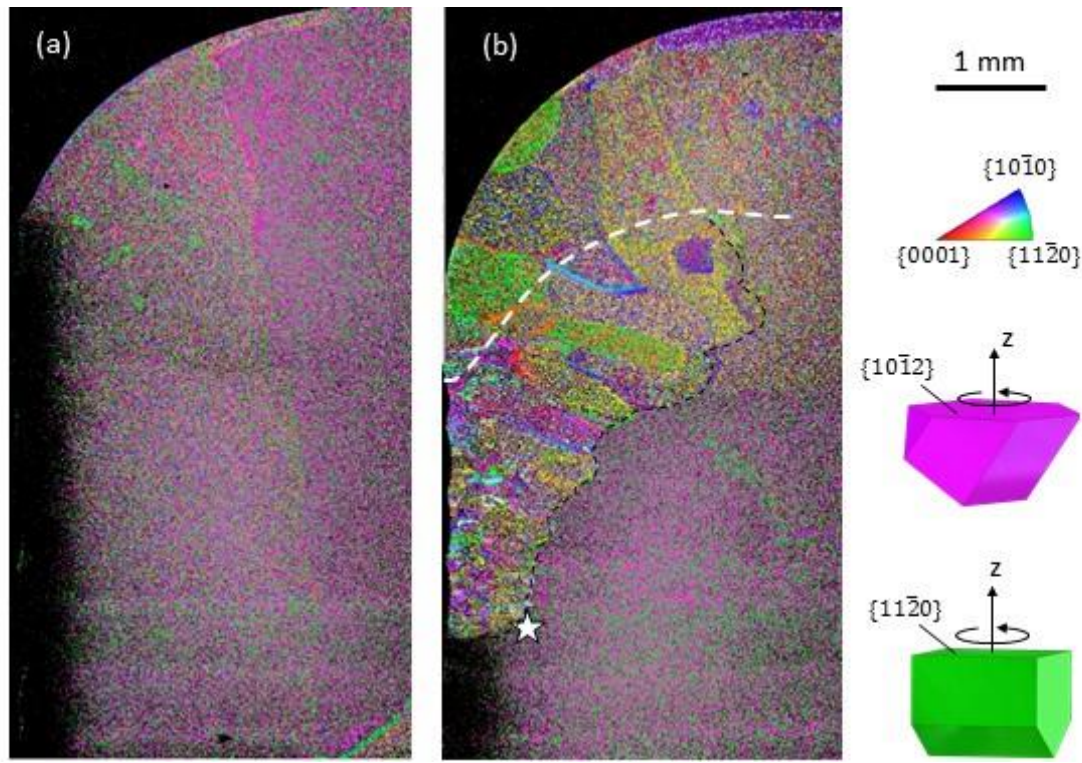

**Figure 12: The α-phase IPF-z map of the top left corner of the wall (a) after MHP and (b) after heat treatment.**

The second most common alignment with this fibre is the green-coloured {1120}$_{hcp}$ plane. No other orientation could be identified in the as-deposited and peened specimen in Figure 12 (a). As shown by Donoghue et al. [15], only these two options can occur according to BOR. However, the variant selection is not only statistical, but also depends on the cooling rate [33] and the β-grain growth [34]. The dark region on the left of the non-heat treated specimen, where MHP was performed could not be indexed. After the heat treatment various α-orientations can be identified within the region of refined β-grains, shown in Figure 12 (b). Each refined β-grain however seemed to contain two dominant α-orientations. The dark region on the left surface was recovered after refinement and indexed by EBSD up to the layer band, the location of which is indicated by the star.

The reconstructed β-phase IPF-z maps of both specimens are shown in Figure 13 (a) and (b). The as-deposited and peened specimen showed a complete alignment of all β-grains with the {001}$_{bcc}$ fibre in z-direction. Not only the β-grain grain size, but also the orientation was influenced dramatically by the heat treatment as shown in Figure 13 (b). Within the refinement boundary, a wide range of β-orientations could be identified without any of them being dominant. The β-grain size close to the layer band appeared very small, while closer to the fusion boundary, the grains became larger, columnar and aligned perpendicular with the fusion boundary.

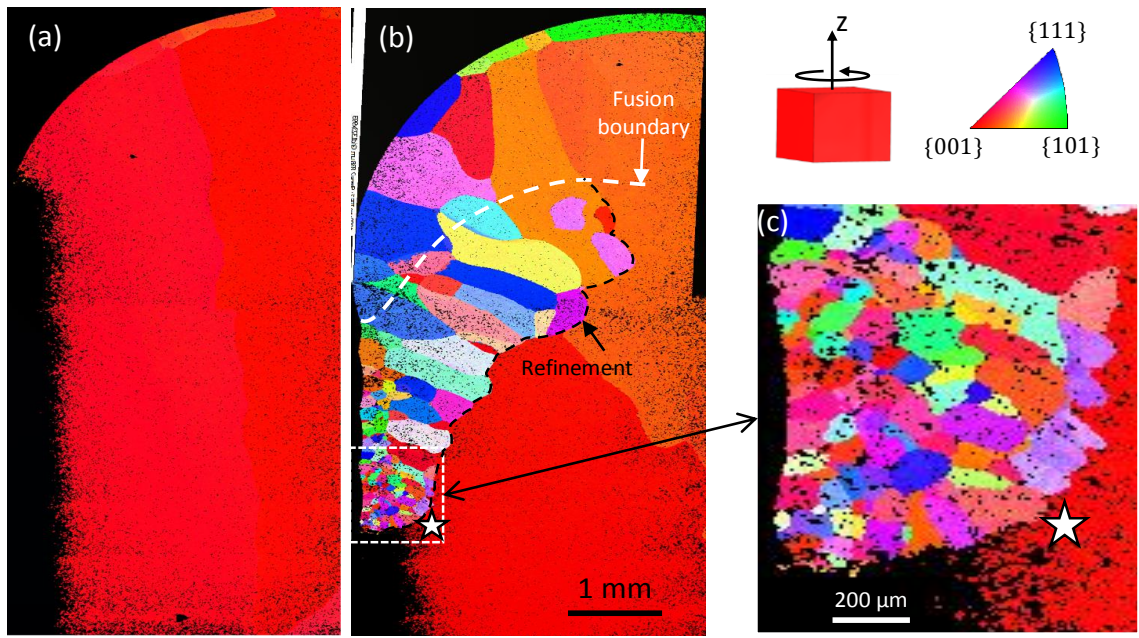

Figure 13: The reconstructed β-phase IPF-z map of the top left corner of the wall (a) after MHP and (b) after heat treatment with (c) a higher magnification showing the refined grain size.

The EBSD strain maps in Figure 14 show qualitatively the amount of plastic strain represented by α-lattice misorientation. Neither specimen contained many lattice imperfections caused by plastic deformation in most of the material, represented by purple coloured α-lattice with a deviation of 1° or less. After MHP strong plastic deformation could be observed between 500 µm and 800 µm depth on the left side of the peened specimen in Figure 14 (a), where the 3° and 4° α-lattice misorientation is indicated with green and yellow. Within 500 µm of the surface the plastic deformation was so severe, that the material could not be indexed, due to extreme plastic deformation. The location of the edge of the wall is therefore represented by the dashed line in Figure 14 (a). The heat treatment recovered the material, which allowed indexing again of the lattices within the refined region. In the high magnification image in Figure 14 (c), at the lower boundary of the refined region, an abrupt transition from plastically deformed to recovered and refined material can be observed. The complete recovery also allowed identification of the locations of the MHP passes on the left surface.

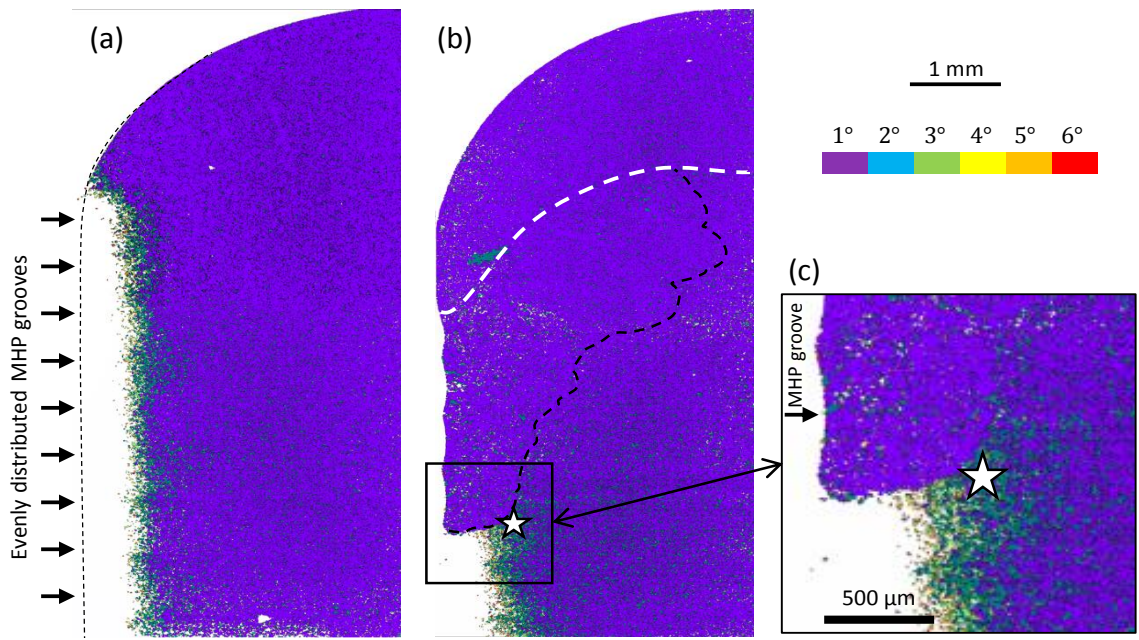

**Figure 14: The EBSD strain map of the top left corner of the wall (a) after MHP and (b) after heat treatment with (c) a higher magnification showing border and depth of plastic deformation and recovery.**

# 4 Discussion

Aluminium stabilises the α phase and vanadium the β phase in titanium alloys. Both elements tend to concentrate in the respective phase they stabilise [35]. Elmer et al. [21], [36] investigated the diffusion kinetics of Al and V in welding of Ti–6Al–4V alloy and found that the V concentration deviates more between both stable room temperature phases, compared to Al. The nominal composition of both elements distribute as follows: The 4 % V can reach 16 % in the β phase and only 1 % in the α phase. The 6 % Al instead can reach 7 % in the α phase and 3 % in the β phase. Therefore the diffusivity of V is the dominant factor during thermal processing and determines (depending on the cooling rate) the delay between the nominal transformation temperature and the temperature at which the transformation actually occurs, as well as the resulting α-plate size [9], [21], [26], [36]. The diffusion distances of V during heating and cooling [9] are comparable to the α-plate thickness in Figure 10 (a-c) of roughly 1 μm and it is also understood to be the driving factor for α and β coarsening [37], [38].

## 4.1 Temperature/time history and resulting microstructure

The logarithmic trendline of the thermocouple readings (Figure 9) allows a precise interpolation between the measurement points, so peak temperatures and thermal histories can be assigned to microstructural key features. The derivative of this trendline provides the temperature gradient at each location as a function of the distance from the top of the deposit (eq. (1)). At the location of the fusion boundary where the columnar grains form (x = 1.5 mm), the gradient is 387 °C/mm, which explains the evolution of the columnar microstructure (Figure 11 (a)) according to Bontha et al. [27].

$$\frac{d}{dx}(T(x) = -580.3\ln(x) + 1832.7) = \frac{-580.3}{x}\left[\frac{°C}{mm}\right] \tag{1}$$

The white band, typical for titanium WAAM walls, was one of the features of most interest, for which the trendline predicts a peak temperature of 825°C. This value compares with 748°C measured by Martina et al. [6] at this location with TIG deposition performed with a slower travel speed, which resulted in a longer exposure allowing more time for diffusion and transformation at similar temperatures. Elmer et al. [9] reported that heating rates of +30 °C/s required 50 °C overheating of Ti–6Al–4V to observe equivalent amount of transformed β-phase compared to equilibrium thermodynamics. The heating rates at thermocouple 3 (almost at the location of the white band) were in fact higher than 500 °C/s (Figure 7(b)), which could have caused the material to require approximately 110°C superheat over the α-dissolution temperature to establish α-plates. It is therefore proposed that the white bands represent the initiated α-dissolution due to the transient thermal field and not just the α-dissolution temperature.

There is a second slightly brighter region below the layer band, which reached between 600°C and 825°C, where the hardness dropped significantly, even though no microstructural changes can be observed compared to the base material using the back-scatter detector of the SEM (Figure 10 (e) and (f)). It is known that diffusion becomes significant at temperatures above 550 °C [39]. This does not yet have an effect on the morphology, but changes the local composition of both phases, which could weaken the material. A modified composition of the α-plates with more Al and less V would also explain the increased resistance against hydrofluoric acid during the etching, causing the slightly brighter appearance

Just above the white line at thermocouple 3, where the material reached 860°C (140°C below the β-transus) and stayed for approximately 2.5 s above the α-dissolution temperature, microstructural changes could be observed using the SEM. Figure 10 (d) shows the formation of α-colonies of many and very thin α-plates (~ 0.3 µm). The morphology of these α-colonies appeared to be related in size and shape to the globular grain morphology in the base material (Figure 4), which did not have a lamellar or bimodal microstructure prior to heating. Here the temperature/time history was sufficient to allow diffusion-based partitioning of the aluminium and vanadium to segregate in the respective phase they stabilise [9]. The result is the establishment of α-plates surrounded by β-shells organised in colonies. The relation of the colony morphology to the parent microstructure suggests that the α-plates arose by diffusion and growth during heating and not during rapid cooling of β-grains.

Thermocouple 2 reached 1060°C and remained for 1.6 s above the β-transus. A Widmanstätten microstructure can be identified at this location in the SEM image in Figure 10 (c). The α-plates here have typical dimensions of WAAM material of approximately 1 μm thickness and 10 μm length [6], but a few thinner α-plates as small as those at thermocouple 3 location persisted, indicated with the white arrows. These remnants suggest that the material did not fully transform into the high temperature β-phase due to the extreme heating rates [9] and insufficient time (1.6 s) and temperature (65 °C) above β-transus [36].

The temperature profile at thermocouple 1, shown in Figure 7 (a), featured two abnormalities. The increase in temperature seems to decrease disproportionally at about 2.3 s, when the material was heated between 1200 °C and the peak temperature of 1260 °C. This is more evident in the heating rate chart in Figure 7 (b). This slight reduction in the heating rate could be due to the latent heat of the phase change when the remaining α transforms to the β phase at this temperature [40]. This observation is in agreement with Elmer et al. [36], who found that approximately 4 s of time is required between 120 °C and 200 °C superheat above the β-transus to complete the full transformation, regardless of the welding parameters. In a later study [21] they found that diffusion kinetics can drastically delay the α→β transformation to higher temperatures at more extreme heating rates closer to the weld pool. In fact, the α→β transformation with the highest heating rate (360 °C/s at 1000 °C) occurred at approximately 1500 °C. Slower heating rates are therefore required for the material to complete the transformation at lower temperatures [21]. The Widmanstätten morphology of the lamellar microstructure at thermocouple 1 suggests a complete α→β transformation, because (unlike thermocouple 2) the microstructure does not seem to be related to the base material morphology.

Figure 15 shows the detailed view during cooling at thermocouple 1, taken from Figure 7 (a) and (c) and provides the precise cooling rates with the corresponding temperature. After reaching the peak temperature at thermocouple 1, the material cooled at approximately 90 °C/s through the β-transus at 5.5 s. The temperature then shows a short plateau-like profile at 780 °C between 10 s and 11 s and the cooling rate at this point decreased from −20 °C/s to −10 °C/s, after which it increased again to −20 °C/s. This suggests that the β→α phase transformation occurs at this point. The transformation temperature matches the continuous cooling diagram reported by Ahmed and Rack [26], which predicted a transformation temperature of 770 °C at a cooling rates of −20 °C/s. Kherrouba 2016 et al. [41] reported the precise temperature-dependent phase fraction during β→α transformation with continuous cooling from the β-phase field and showed that the transformation is half completed at 860 °C at a cooling rate of −50 °C/s. Finally Bermingham et al. [19] observed this transformation plateau at 820 °C with tungsten inert gas deposition. The reason for the lower transformation temperature in the present work using a plasma heat source (plateau at 780 °C) is the much faster travel speed (7 mm/s vs. 0.8 mm/s), less heat input and associated faster cooling rates than deposition with a tungsten inert gas welding torch [19]. Since this is a microstructural effect, thermal simulations fail to reproduce it.

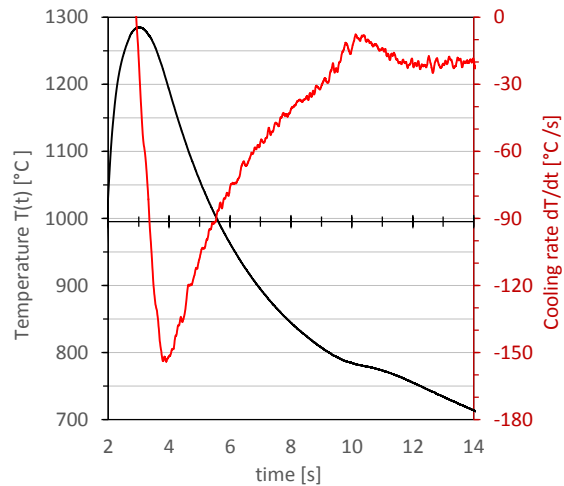

Figure 15: Transient cooling at thermocouple T1 showing isothermal hold during β → α transformation.

## 4.2 Grain refinement after MHP

The geometry of the specimen used to identify the thermal history was identical to the deposited WAAM wall, on which inter-pass MHP was applied. This allows the transfer of the thermal profile to the refinement experiment to some extent. The purpose of the MHP experiment was to identify the relationship between plastic strain and thermal history, which affected the evolution of the β-grains. It was found that the region above the last layer band is the location with the lowest temperature that allows refinement, shown in Figure 11 (b). The peak temperature at this point was supposedly 825 °C based on the earlier study and the material was kept above 748 °C for approximately for 3 s. The β-grain size in this region, as shown in Figure 13 (b), is homogeneously distributed and the smallest in the refined field with grains between 20 μm and 150 μm in size. Within 1 mm of the fusion boundary at the surface of the peening contact, the β grains become larger. As MHP produced the same plastic deformation on the entire side surface, it can be concluded that the temperature exceeded the β-transus to an extent that allowed full α→β transformation and immediate subsequent β-grain growth. As reported in Pilchak et al. [38] β-grain growth occurs a few seconds after the material is brought to temperatures of just 15 °C above the static β-transus, depending on the initial β-grain size. Repeated α→β transformation on the other hand is not understood to lead to grain growth, if the initial grain size is already coarse [33]. The location of the plastic strain (and not the magnitude) and therefore the thermal history play a larger role on grain refinement than previously expected [6], where β-grain growth was not considered after refinement.

Something that has not been considered in this study is subsequent refinement of the refined microstructure as proposed by Donoghue et al. [15]. This can occur providing the deformation induced by the roller is sufficiently deep to penetrate into the already refined microstructure. This explains why the β-grains refined by inter-pass MHP close to the layer band and with severe plastic deformation did not produce a finer β-grain size compared to inter-pass rolling [30]. This finding also explains why smaller prior β-grain sizes can be achieved by increasing the rolling load: the strain is induced deeper within the material allowing refinement of the already refined material; and the lower temperatures further from the heat source impede subsequent growth, rather than more strain than producing smaller grains.

A second finding is the suspicious depth of the refinement region near the fusion boundary in the present work (Figure 11 and Figure 13), which reaches 3 mm from the peened surface into the material. The EBSD strain map after peening in Figure 14 (a) suggests that plastic strain was limited to 1 mm from the surface. The region with the large columnar grains can therefore not be evidence of refinement. Instead these grains started to grow from the surface at the top-most peening pass, where refinement and reorientation started. The appearance of the β-grains strongly suggest that they grew from the side of the wall towards the centre of the deposit before the region where they nucleated was completely remelted and the high temperatures in this region allowed this growth to occur which is not observed further from the heat source. The molten pool then solidified epitaxially on these medium sized columnar β-grains with a new orientation based on previous epitaxial growth inside the bulk, as the heat source approached.

The findings in the present work and in-situ EBSD measurements of Ti–6Al–4V during heating to β-transformation [28] also offer another explanation why inter-pass cold work refines the α-plates: it was observed that the α-plate size depends on the cooling rate through the transformation temperature range (Figure 10 (a) and (b)). However, it was also observed that material that has never completed the α→β transformation can still feature α-plates, as shown in Figure 10 (d). The unit cells of the α-plates and of the surrounding retained β-shells are aligned according to BOR. Subsequent heat cycles on the same material, as they occur in WAAM, can cause further α-plate growth, without being hindered by the growth of the retained β-phase. Rolling however probably twins the β-unit cell [28] and thereby resets the orientation memory effect [34], when the β-grains regrow. Heating of twinned β-grains promotes the growth in a different orientation, which imposes the new β-orientation on the α-phase [28]. The α-grains are figuratively being "consumed" by the growing twinned β-grains, they lose their texture memory and start to grow from scratch with a BOR variant of the new β orientation. The resulting α-plate's thickness is therefore as small, as if the material has only seen one thermal cycle and no re-heating and explains why rolling and peening not only reduce the β-grains, but also the α grains.

# 5    Conclusion

1. The thermal history in the HAZ in WAAM was resolved in detail for one set of parameters. The peak temperatures, heating rates and cooling rates, as well as temperature gradients allow detailed discussion of the resulting microstructure.

2. The temperature gradient during solidification of 386 °C/mm is responsible for the formation of the columnar microstructure. For an equiaxed 100 °C/mm would have been required.

3. 60 °C of superheat above the β-transus and 1.6 s was not sufficient for α→β transformation (T2). It required approximately 200 °C superheat and 3.8 s due to finite diffusion kinetics at the high heating rate of over 1000 °C/s for full transformation (T1). Cooling rates of between 20 and 90 °C/s delayed the β→α transformation to 780 °C.

4. Lamellar microstructure can grow during heating above 860 °C of non-lamellar microstructure, without cooling from the β-phase field. It is likely that the lamellar microstructure can also form at lower temperatures, if the heating and cooling rates were slower.

5. In this particular experiment, the peak temperature at the white band in the microstructure, which is also observed in material produced with WAAM was approximately 825°C – well below the β transus temperature but above the α dissolution temperature of 748°C.

6. Significant softening of the wrought material occurred below the layer band for peak temperatures between 550°C and 748°C. This degree of softening does not occur in material produced with WAAM.

7. Microstructural features are caused by temperature/time history and cannot be labelled as a certain peak temperature achieved, as the diffusion kinetics also affect the transformation.

8. Grain refinement can be achieved just above the last layer band, which is where the smallest grains occur are produced. Therefore, plastic deformation should be induced ideally just above the region that will become the white layer band during the subsequent deposition to achieve the greatest grain refinement.

9. Increasing the plastic strain, decreases the β-grain size after refinement. However, once a certain amount of strain is induced, the thermal profile becomes the limiting factor for the smallest β-grain size that can be achieved, as very small grains can grow after they have been refined if the temperature during the thermal cycle is high. Grain refinement near the fusion zone can result in β-grain growth within material that hasn't been plastically worked.